\documentclass[compsoc, conference, a4paper, 10pt, times]{IEEEtran}

\usepackage{cite}
\usepackage{amsmath,amssymb,amsfonts}
\usepackage{algorithmic}
\usepackage{graphicx}
\usepackage{textcomp}
\usepackage[table,xcdraw]{xcolor}
\usepackage{booktabs}
\usepackage{multirow}
\usepackage[hidelinks]{hyperref}

\usepackage[nolist]{acronym}
\usepackage{multirow}
\usepackage{pgfplots}
\pgfplotsset{compat=1.14}
\usepgfplotslibrary{statistics}
\usepackage{filecontents}

\usepackage{booktabs}
\usepackage{multirow}
\usepackage{todonotes}

\usepackage{diagbox}
\usepackage{listings}
\usepackage{mwe}
\usepackage{orcidlink}

\makeatletter
\def\ps@IEEEtitlepagestyle{%
	\def\@oddfoot{\mycopyrightnotice}%
	\def\@evenfoot{}%
}
\def\mycopyrightnotice{%
	{\footnotesize 
		\begin{minipage}{\textwidth}
		\textcopyright 2023 IEEE.  Personal use of this material is permitted.  Permission from IEEE must be obtained for all other uses, in any current or future media, including reprinting/republishing this material for advertising or promotional purposes, creating new collective works, for resale or redistribution to servers or lists, or reuse of any copyrighted component of this work in other works.\\
		\href{https://doi.org/10.1109/EuroSPW59978.2023.00049}{DOI 10.1109/EuroSPW59978.2023.00049}
		\end{minipage}
		}
	\gdef\mycopyrightnotice{}
}

\begin{document}

\newcommand{\ourtitle}{\textsc{D-GATE}}
\title{\ourtitle{}: Decentralized Geolocation and Time Enforcement for Usage Control}


\author{\IEEEauthorblockN{1\textsuperscript{st} Hendrik Meyer zum Felde\orcidlink{0000-0002-5837-8730}}
\IEEEauthorblockA{\textit{Fraunhofer AISEC} \\
Garching near Munich, Germany \\
hendrik.meyerzumfelde@aisec.fraunhofer.de}
\and
\IEEEauthorblockN{2\textsuperscript{nd} Jean-Luc Reding\orcidlink{0000-0003-3176-3388}}
\IEEEauthorblockA{\textit{Fraunhofer AISEC} \\
Garching near Munich, Germany \\
jean-luc.reding@aisec.fraunhofer.de}
\and
\IEEEauthorblockN{3\textsuperscript{rd} Michael Lux\orcidlink{0009-0008-7533-2110}}
\IEEEauthorblockA{\textit{Fraunhofer AISEC} \\
Garching near Munich, Germany \\
michael.lux@aisec.fraunhofer.de}
}

\maketitle


\begin{abstract}
  In the context of cloud environments, data providers entrust their data to data consumers in order to allow further computing on their own IT infrastructure. 
Usage control measures allow the data provider to restrict the usage of its data even on the data consumer's system.
Two of these restrictions can be the geographic location and time limitations. 
Current solutions that could be used to enforce such constraints can be easily manipulated. 
These include solutions based on the system time, organizational agreements, GPS-based techniques or simple delay measurements to derive the distance to known reference servers. 

With D-GATE, we propose a reliable solution that uses trusted execution environments and relies on a decentralized mesh of reference nodes, so-called GeoClients.
Here, participants periodically measure the lowest network delay to each other to geolocate themselves.
For data providers, it is thus possible to technically attest usage control with time and geolocation constraints without depending on centralized reference systems.
\end{abstract}


\begin{acronym}[ECU]
  \acro{ias}[IAS]{Intel Attestation Service}    
  \acro{dh}[DH]{Diffie-Hellman}
  \acro{ecdsa}[ECDSA]{Elliptic Curve Digital Signature Algorithm}     
  \acro{ecdh}[ECDH]{Elliptic Curve Diffie-Hellman} 
  \acro{tee}[TEE]{Trusted Execution Environment}
  \acroplural{tee}[TEEs]{Trusted Execution Environments}
  \acro{tpm}[TPM]{Trusted Platform Module}
  \acroplural{tpm}[TPMs]{Trusted Platform Modules}
  \acro{cpu}[CPU]{Central Processing Unit}
  \acroplural{cpu}[CPUs]{Central Processing Units}
  \acro{eds}[EDS]{Enclave Definition Service}
  \acro{uas}[UAS]{User Authentication Service}
  \acro{eas}[EAS]{Enclave Attestation Service}
  \acro{me}[ME]{Management Enclave}
  \acro{doe}[DOE]{Data Operation Enclave}
  \acroplural{doe}[DOEs]{Data Operation Enclaves}
  \acro{as}[AS]{Attestation Service}
  \acro{ed}[ED]{Enclave Definition}
  \acroplural{ed}[EDs]{Enclave Definitions}
  \acro{e2e}[Enc2Enc]{Enclave to Enclave}
  \acro{pep}[PEP]{policy enforcement point}
  \acro{r2e}[Rem2Enc]{Remote to Enclave}
  \acro{edmm}[EDMM]{Enclave Dynamic Memory Management}
  \acro{epc}[EPC]{Enclave Page Cache}
  \acro{ac}[AC]{Attestation Command}
  \acro{tr}[TR]{Timestamp Request}
  \acroplural{tr}[TRs]{Timestamp Requests}
  \acro{st}[ST]{Signed Timestamp}
  \acroplural{st}[STs]{Signed Timestamps}
  \acro{ar}[AR]{Attestation Report}
  \acro{dr}[DR]{Data Request}
  \acro{os}[OS]{Operating System}
  \acroplural{dr}[DRs]{Data Requests}
  \acro{tcb}[TCB]{Trusted Computing Base}
  \acro{tpm}[TPM]{Trusted Platform Module}
  \acroplural{tpm}[TPMs]{Trusted Platform Modules}
  \acro{poc}[PoC]{Proof-of-Concept}
  \acro{pep}[PEP]{policy enforcement point}
  \acro{mecl}[MECL]{Multi Enclave based Code from List}
  \acro{mect}[MECT]{Multi Enclave based Code from Template}
  \acro{seif}[SEIF]{Single Enclave based Interpretation Framework}
  \acro{pki}[PKI]{Public Key Infrastructure}   
  \acro{cl}[CL]{Client List} 
  \acro{gr}[GR]{Geolocation Report} 
  \acro{mr}[MR]{Measurement Request}   
  \acro{vm}[VM]{Virtual Machine} 
  \acro{sm}[SM]{Signed Measurements} 
  \acro{sgx}[SGX]{Software-Guard-Extensions}   
  \acro{sev}[SEV]{Secure Encrypted Virtualization}
  \acro{sev-snp}[SEV SNP]{Secure Encrypted Virtualization with Secure Nested Paging}
\end{acronym}

\section{Introduction} \label{introduction}

In many different applications in industry, science and business, data is shared to create new values from it.
Data providers often want to share their data with data consumers and allow them to process this data in the data consumer's own IT infrastructure.
In this case, a data provider must implement and enforce measures of control on the consumer's side to strictly limit the data processing to its initial purpose and to remain sovereign over shared data.
Simply trusting the consuming peer is not always an acceptable option. 

The need to control data sovereignly after its release is a new common use case for data management which has evolved within the last decade.
To address this use case the concept of usage control was formalized in 2004 by Park and Sandhu \cite{park2004uconabc}.
Usage control pairs common access control mechanisms with an additional enforcement of usage constraints on the data consumer's side.
This formalization was called the ``UCON ABC'' which models access control using authorizations, obligations, and conditions.

Two important and relevant conditions to enforce in this context are \textit{geolocation} and  \textit{time} constraints, which are the focus of this work. 
Restrictions for these properties may be required from business models, such as different pricing sets of data for specific regions, or temporary limited collaborations, where the data must not be usable after a certain time.
Another very strong reason for such limitations are legal regulations. 
For example, the General Data Protection Regulation (GDPR) of the EU only allows the processing of personal related data on machines located outside the EU under some further conditions (e.g., adequacy decision or appropriate safeguards)\cite[Article 44]{GDPR}.

Current approaches to remotely enforce or attest geolocation and system time can be manipulated too easily due to the following reasons. 
Concerning solutions using external sources, such as timeservers or geolocation certificates, have the security drawback of having to trust the accuracy and integrity of a specific service.
Additionally, the system time can easily be changed, and virtual machines can be set back to an earlier snapshot of their execution or temporarily paused. 
For geolocation, GPS signals can be spoofed, and IP-addresses can be tunneled and therefore both do not provide the required reliability. 
Geolocation techniques based on network topology are too inaccurate and highly rely on the integrity of chosen reference nodes \cite{gill2010dude,fotouhi2015plag,jinxia2016ip}.
A geolocation based on measurements of network delay seems a bit more promising as minimal network delays cannot be forged or spoofed. 
There are already solutions with delay measurements which we want to make more secure.
However, it is necessary that the physical location of servers used as references is also verified before using it for the location determination for a client.
All in all, better solutions with more precision and less security trade-offs need to be found.

Another problem is the fact that the enforcement of usage constraints requires software on the data consumer's side which must be known to work in a trustworthy and non-modified state. 
The trustworthy initial state is usually remotely attested via hash measurements by secure hardware modules which serve as trust anchor.
Examples of such hardware modules are \acp{tpm}, which are secure co-processors typically attached to motherboards, or \acp{tee} which allow a protected execution of software and are typically used in cloud-environments.
\acp{tpm} and \acp{tee} can consume data and produce hashing fingerprints and can contain and protect cryptographic key material. 
Most \acp{tee} provide additional security primitives such as confidentiality and integrity protection for processes or VMs which is a mandatory  feature for enforcing usage control.
Otherwise, the usage control could be manipulated or data directly extracted.
Therefore, transferring the enforcement of time and location constraints into a \ac{tee} enhances the security and prevents manipulations of the software.

We need to enable usage control with the enforcement of time and geolocation constraints to fulfill current real world demands, but the following problems are at hand. 
First, \acp{tee} are not capable to natively attest geolocation or internal time settings. 
Second, \acp{tee} add a layer of imprecision for measurements of network delays. 
Third, the aforementioned native attestation techniques only cover an attestation directly after a system's initialization and additional solutions for continuous attestation are required.
The system we introduce addresses the mentioned issues by using a massive mesh of TEE-protected clients which regularly determine the geolocation and time settings of nearby clients using network delay-based measurements for geolocation. 

We provide the following contributions:
\begin{itemize}
  \item We present \ourtitle{}, a design for \acp{tee} which adds the enforcement of time and geolocation constraints to usage control during runtime and makes the enforcement attestable to a third party.
  \item We introduce our TEE-agnostic design together with a delay measurement implementation for \ourtitle{} which can be executed using commonly known TEEs such as Intel SGX and AMD SEV. \footnote{The delay measurement implementation and the testing environments for Intel SGX and AMD SEV can be found at \url{https://github.com/Fraunhofer-AISEC/dgate}.}
  \item We provide a detailed precision evaluation of the geolocation constraints using \ourtitle{} in both, an Intel SGX and an AMD SEV environment.
  \item We introduce a methodology to detect shifts of location of reference systems.
\end{itemize}

\section{Related Work}\label{relatedwork}
Various methods have been proposed to determine the location of a computer using techniques based on measuring network delay. 
Nevertheless, these papers have not set the focus on security and reliability and remain insecure, which we aim to address with our work.

In 2001, Padmanabhan and Subramanian analyzed location techniques using IP and network-based delay measurements to known reference nodes \cite{padmanabhan2001investigation}.
In 2006, Gueye et al. further extended and improved the idea of GeoPing by increasing the amount of usable reference points for delay measurements \cite{gueye2006constraint}. 
Also in 2006, Katz-Bessett et al. proposed a solution \cite{katz2006towards} which does not consider landmarks as reference points, but instead network topology.
Both works still lack security.

In 2010, Gill et al. presented attacks on the above-mentioned techniques and provided their evaluation \cite{gill2010dude}.
The attacks were based on a threat model where adversarial clients try to forge a geolocation by delaying certain network measurements on purpose, which is similar to the threat model assumed in this paper. 
They concluded, that an attacker could increase the delay from selected landmarks and shift the calculated location of the target with the cost of a trade-off between accuracy and detectability.
Also in 2010, Eriksson et al. proposed an improvement to network-based geolocation via machine learning techniques \cite{eriksson2010learning}. 

In 2015, Fotouhi et al. aimed to geolocate the datacenters of cloud providers instead of landmarks and slightly improve the efficiency of constraint-based network geolocation \cite{fotouhi2015plag}.
In 2016, Jinxia et al. introduced a precision improvement based on probability measurements and took into consideration the different characteristics of the network topology in China which has a lot less correlation between distance and network delay compared to Europe or America \cite{jinxia2016ip}.
They obtained precisions of 4km of error as median for 996 chosen reference landmarks. For 40\% of the targets they achieved an accuracy of 3km and for 13\% of the targets the accuracy was larger than 15km \cite[Sec.~VI-B]{jinxia2016ip}.
Also in 2017, Stephanow et al. developed techniques to continuously locate the position of cloud providers despite frequent network changes \cite{stephanow2017continuous}.
They used an adaptive learning scheme with experimental results for 14 different cloud service components within the Amazon Web Services Global Infrastructure. 

In summary, the characteristics of geolocation techniques based on network delay have already been extensively studied. 
However, the mentioned works mostly cover and analyze scenarios, where someone can geolocate her own system on her own behalf.
The related work proposed so far does not cover the security aspects that become relevant when the geolocation of a data processing entity has to be proven to another party, i.e. a data provider.
Proving the geolocation to someone else and binding this piece of information to a process executed inside of a \ac{tee} is an important missing link addressed in this paper.
Additionally, we aim to be as independent as possible concerning third parties, such as organizations or governments which provide trusted reference systems in the related work mentioned. 

Besides the aforementioned purely network-based techniques for geolocation, numerous \ac{tee}-based solutions, partly in combination with GPS, have been proposed.
In 2013, Gondree and Peterson provided and evaluated a framework for binding data of cloud providers to a certain location using constraint-based network geolocation techniques \cite{gondree2013geolocation}.
Consequently in 2014, Fu et al. proposed a protocol based on the previous work to pair network-based geolocation with a \ac{tpm} to bind stored cloud data to a machine and its location within a trusted execution environment\cite{fu2015trusted}.

In 2015, Park et al. proposed another \ac{tee}-based approach using a hypervisor-based attestation technique, which uses GPS coordinates in combination with a TPM as trust anchor.
However, they assumed, that the GPS device could not be forged.
As a worst-case scenario, the transmitted GPS data could easily be spoofed without exchanging the receiving GPS module. 
For this reason, our work does not take into account any GPS-based solutions, although this technology would provide very high accuracy and could be easily integrated into a certification mechanism.

None of the related work mentioned so far covers the impact on network delay measurements executed by processes running inside of a \ac{tee}, which we focus on in our work.
Also, our work deals with techniques which provide definite proof for geolocation which cannot be manipulated, as in the case of GPS-based approaches.
\section{Background} \label{IntroSGXLabel}

Numerous \acp{tee} are available these days \cite{jauernig2020trusted} and even though our design can be used independently of a certain \ac{tee}, some background knowledge on TEEs is required to understand the reasoning within this work.
On the one hand, there exist \textit{\ac{vm}}-based \acp{tee}, which protect, execute, and attest software stacks involving a VM, such as AMD \ac{sev} \cite{kaplan2016amd}.
AMD SEV protects its VMs memory by encrypting each VM individually using a secure co-processor.
The performance overhead is mostly due to the required AES encryption and decryption of required memory pages at usage.

On the other hand, there exist \text{process}-based \acp{tee} which protect, execute and attest processes only, such as Intel \ac{sgx} \cite{johnson2016intel}.
SGX provides so-called secure enclaves which provide primitives to ensure the security goals integrity protection and confidentiality \cite{costan2016intel}. 
In this TEE, the execution of a process is separated from the rest of the system via hardware mechanisms and encryption.
However, switching into an enclave causes the CPU to make a costly context switch, that requires the system to decrypt and reload the last state of the protected process, which is to be resumed.
In both technologies the threat model addresses scenarios where either a malicious hypervisor or host \ac{os} attempts to break into the protected executions of the \acp{tee}.

Regarding delay measurements, the processing time required to execute a geolocation protocol is crucial and should fluctuate as little as possible.
In general context switches within \acp{tee} produce a performance overhead which has an impact on delay measurements.
This is the case for context switches from an untrusted host \ac{os} to either a protected guest system or into a protected process.
Once the process needs to reach system services, such as network components, another context switch back to the untrusted \ac{os} is required, whereas VMs in AMD SEV can directly access network hardware etc.

The original intend behind SGX was to outsource and protect only important parts of an application and carefully whitelist the required interfaces to the untrusted \ac{os} to reduce the attack surface.
However, in industry the need to launch completely unmodified applications was at hand due to costly otherwise required code adaptions, as it would have been required for our testing implementation.
This need was met by frameworks like ego-dev \cite{egodev}, Gramine-SGX (formerly known as Graphene-SGX) \cite{tsai2017graphene,tsai2014cooperation} and Scone \cite{arnautov2016scone} which allow to compile non-modified go applications to a framework-specific binary that can be executed in SGX.
The trade-off for the simplified execution is an increased \ac{tcb} because the framework's code which becomes part of an enclave must be attested and trusted, also.
AMD SEV instead, does not require additional frameworks, but requires a VM to run programs.

\section{Design}\label{design}

\begin{figure}
    \begin{center}
      \includegraphics[width=1\linewidth]{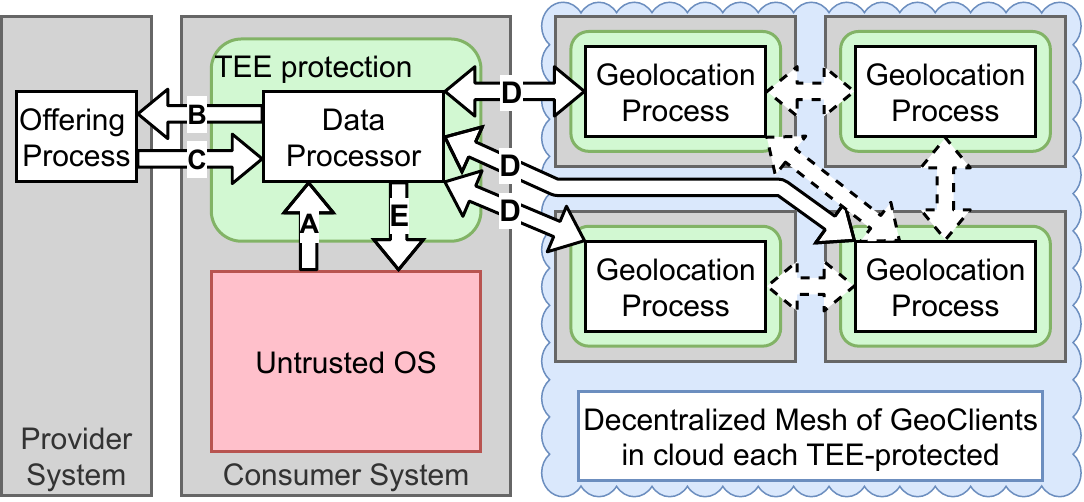}
      \caption{Overview of geolocation process with context and interaction overview of involved Data Provider, Data Consumer's \ac{tee} and \ac{tee}-protected geolocation processes (green round) which interact with the cloud mesh of GeoClients (blue cloud) and the Data Consumer's Untrusted OS (red rectangle).}
      \label{figure:ContextOverview}
  \end{center}
\end{figure}

\begin{figure}
    \begin{center}
      \includegraphics[width=1\linewidth]{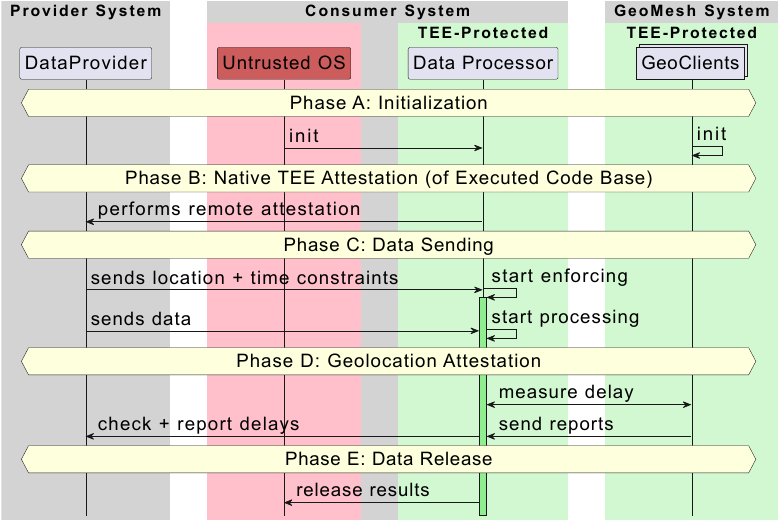}
      \caption{ 
    Simplified sequence diagram showing an overview of the interactions of the involved Data Provider, Data Consumer's Untrusted \ac{os}, the \ac{tee}-protected data processor, and the \ac{tee}-protected mesh of GeoClients.
    }
      \label{figure:SequenceAttestation0}
  \end{center}
\end{figure}

To enforce time and location constraints in usage control, our design involves three different remote parties consisting of the Data Provider (wants to share its data), the Data Consumer (uses the data of the data provider), and a mesh of GeoClients which are placed in the cloud and used for decentralized geolocation. 
The Data Consumer consists of a \ac{tee} and an Untrusted \ac{os}.
The Data Consumer's \ac{tee} and each GeoClient's \ac{tee} must be trusted.
Figure \ref{figure:ContextOverview} provides an overview of the architecture with involved participants and key elements for the proposed concept.
Figure \ref{figure:SequenceAttestation0} provides a simplified overview of the sequence of steps.

The core idea of \ourtitle{} is a mesh of \ac{tee}-protected GeoClients in which each participant periodically measures the network distance to each other and provides these measured relative delays upon request. 
This massive use of decentralized reference nodes prohibits the adversarial spoofing of network locations.
The measurements allow to derive a geolocation among all other GeoClients, which must comply with the limiting rules for data processing given by the data provider.
The involved phases of both figures are introduced in the following sections.

\textbf{Phase A - Initialization:}
In the beginning, the consumer's Untrusted OS must initialize a known DataProcessor inside of a \ac{tee}. 
Additionally, the set of GeoClients which form our mesh of protected GeoClients must be initialized, also.
This means that existing GeoClients start to keep track of minimal delays to each other and at least one GeoClient must be associated with a known location by the Data Provider.

\textbf{Phase B - Native \ac{tee} Attestation:}
Consequently, the Data Consumer's side must perform a remote attestation of its Data Processor's executed code base to the Data Provider.
The Data Processor is executed inside a \ac{tee} and is therefore able to use the \ac{tee}'s natively provided attestation techniques.
The native attestation guarantees that the Data Processor's code base, which must be known to both provider and consumer, is in a non-modified, trustworthy state at initialization.

\textbf{Phase C - Data Sending:}
If the native attestation succeeds, the Data Provider sends usage constraints (for time and location) and the corresponding raw data to the Data Processor via encrypted communication, which is established during attestation.
The attested Data Processor must be designed to continuously enforce the constraints given by the Data Provider.
After the enforcement, the processing of data may begin confidentially inside the \ac{tee}. 
However, exporting results is prohibited and requires a recently acquired additional attestation of geolocation and time enforcement. 

\textbf{Phase D - Geolocation Attestation:}
\ourtitle{} demands a \textit{geolocation attestation} before the release of calculation results to enforce usage control.
The Data Processor must contact a set of GeoClients chosen by the Data Provider and gather their signed geolocation reports which must contain the following information.
First, the direct network delay measured between Data Processor and GeoClient. 
Second, a list containing a predefined number of the closest geolocation delays to neighbors of other GeoClients.
Third, the GeoClient's current system time. 
The protocol expects asymmetric cryptography and the corresponding \ac{pki} to be applied for the GeoClient's digital signatures.  
These measured delays of strategically chosen GeoClients allow the Data Provider to perform a decentralized check of geolocation and system time within the Data Processor.

\textbf{Phase E - Data Release:}
The Data Processor is implemented in such a way that results can only be passed out to the Untrusted \ac{os} under the control of the consumer if the Data Provider's time and geolocation constraints are fulfilled.
These are the basic phases involved in \ourtitle{}.

\subsection{Detailed Sequence Diagrams}
\label{label:sequencediagramsdetailed}

\begin{figure}
  \begin{center}
    \includegraphics[width=1\linewidth]{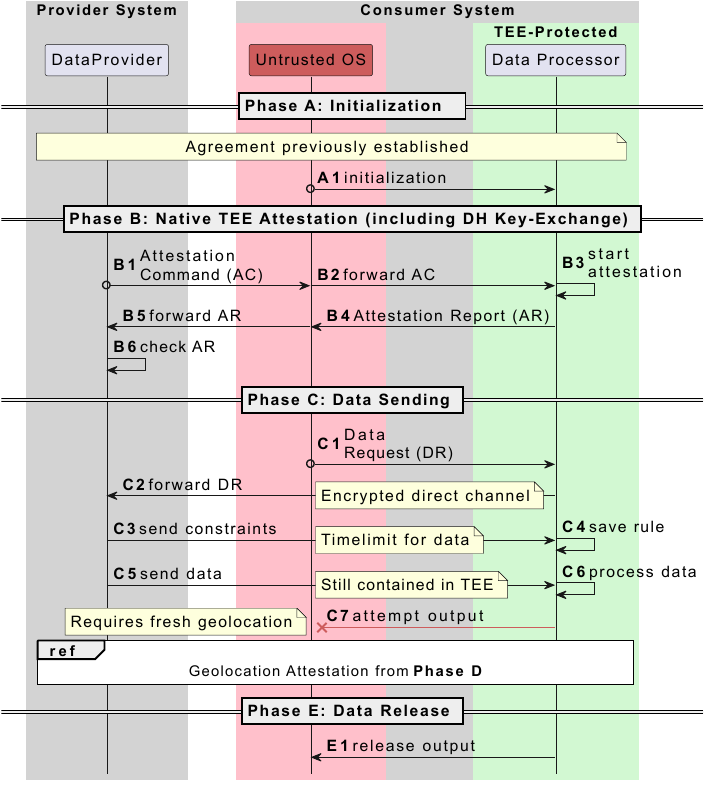}
    \caption{ 
    Sequence diagram showing the interactions of the involved Data Provider, Data Consumer's Untrusted \ac{os} and the \ac{tee}-protected data processor. 
    Focus is set on procedures for initialization of encrypted communication, attestation of the \ac{tee}-protected data processor to the data provider and the communication between parties.
    }
    \label{figure:SequenceAttestation1}
  \end{center}
\end{figure}

\begin{figure}
  \begin{center}
    \includegraphics[width=1\linewidth]{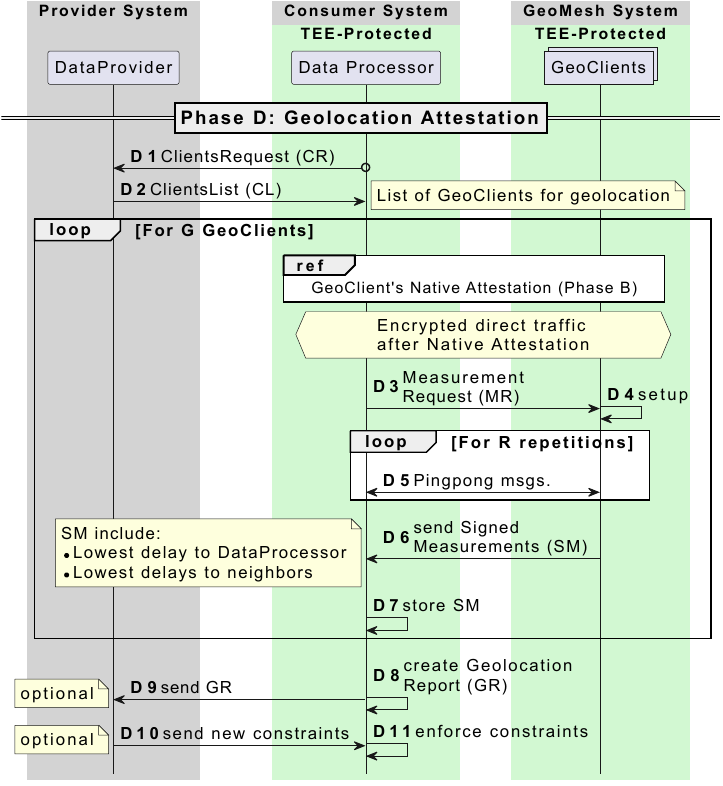}
    \caption{
      Sequence diagram showing the interactions of the involved data provider, \ac{tee}-protected data processor and \ac{tee}-protected GeoClients. 
      Focus is set on the data flow during transmission from the data provider to the data processor and the blocked release of the processing results which require a fresh geolocation.}
    \label{figure:SequenceAttestation2}
  \end{center}
\end{figure}

For better comprehension, this section provides more detailed sequence diagrams of the involved steps for the attestation as can be seen in Figures \ref{figure:SequenceAttestation1} and \ref{figure:SequenceAttestation2} which involve the same parties as in the context view from the previous figures.
Beginnings of interactions are marked with a circle and arrows define the successor of each action.
The higher level of detail is needed to understand the upcoming analysis and the impact of \acp{tee} on delay measurements which is evaluated later.

In \textbf{Phase A}, the protocol assumes that a previously negotiated agreement between Data Provider and Data Consumer exists.
The Untrusted \ac{os} starts by initializing an instance of a known Data Processor inside of an attestable \ac{tee}. 

In \textbf{Phase B} the Data Provider starts by sending an \ac{ac} to the Untrusted \ac{os} (B1) which it forwards right into the \ac{tee}-protected Data Processor (B2).
The \ac{ac} also contains a nonce and material for establishing a key exchange similar to TLS.
Afterwards, the Data Processor starts the attestation by triggering \ac{tee}-specific measurements of the initial code base for the process which digests the nonce and creates an \acf{ar} which is bound to the ongoing key exchange via digital signatures (B3).
The \ac{ar} is encrypted and sent to the Untrusted OS (B4) which forwards the \ac{ar} to the Data Provider (B5). 
The Data Provider must verify the \ac{ar} (B6) and only if it is valid the protocol proceeds.
Otherwise, the system is immediately considered to be compromised. 
However, these failure procedures were left out of the figures for the sake of simplified visualization.  
The native attestation created a TLS key exchange which allows to use encryption for further communication between the Data Provider and the Data Processor.
Therefore, the subsequent elements of the diagram are simplified even though all communication is technically still passed through the Untrusted OS.

In \textbf{Phase C} the Data Consumer creates a \ac{dr} via the Untrusted \ac{os} and sends it into the previously correctly attested DataProcessor (C1), which forwards it further to the Data Provider (C2).
The \ac{dr} must contain a credential which links the data consumer's subscription agreement to the DataProcessor.
If the \ac{dr} is validated correctly at the Data Provider, it sends a set of constraints to enforce for upcoming data (C3) which must be saved and immediately and continuously enforced at the DataProcessor (C4).
Constraints would be time and geolocation limits, which at this point the Data Provider can trust to be enforced on the consumer's side.
Consequently, the Data Provider sends the requested data to the DataProcessor (C5) where immediate preprocessing is starting (C6).
However, as soon as results or intermediate outputs are requested by the data consumer via the Untrusted \ac{os}, the DataProcessor must provide a fresh and valid Geolocation Attestation, denoted as \textbf{Phase D}.
Otherwise, in \textbf{Phase E}, no results can be received from the DataProcessor (E1).

In \textbf{Phase D} the details of the Geolocation Attestation are shown, as illustrated in Figure \ref{figure:SequenceAttestation2}.
Here, the Untrusted OS does not play a role anymore due to end-to-end encryption. 
Instead the messages involve the three parties Data Provider, the consumer's DataProcessor and a large, distributed mesh system of GeoClients which are each \ac{tee}-protected.
First, the DataProcessor requests a list of GeoClients to contact from the DataProvider (D1) which has the aim to make it infeasible for a malicious set of GeoClients to form an evil cooperation and forge timing delays.
The Data Provider replies with the corresponding \ac{cl} (D2). 

Each of the G numbers of GeoClients must natively attest themselves to the DataProcessor as explained in Phase B which additionally establishes an encrypted end-to-end communication channel.
Optionally, if the DataProcessor itself should serve as additional reference node in the GeoMesh a mutual native attestation instead of a single-sided would be required.
Afterwards, the DataProcessor sends a \ac{mr} to the currently addressed GeoClient (D3) which triggers a setup and preparation at the GeoClient (D4).
Consequently, the GeoClient launches a ping pong message exchange which aims to find the lowest network delay with a configurable number of R repetitions (D5).
Next, the GeoClient provides the following pieces of information integrity protected with a digital signature in the \ac{sm} (D6). 
The \ac{sm} contains the lowest measured delay, a nonce and a list of recently measured GeoClients with lowest delays, which each GeoClient must periodically update. 
At the end of the loop for each GeoClient, the \ac{sm} is stored locally in the DataProcessor (D7).

Finally, all the gathered sets of \ac{sm} are combined in a \ac{gr} (D8) and optionally sent to the Data Provider to provide either insights, additional remote control, or an adjustment of the constraints (D9-D10).
These steps are optional because the Data Provider's given constraints and pre-programmed decisions must be automated and directly enforced by the validated DataProcessor in any case (D11).
The Data Provider can now be assured that \ac{gr} contains integrity protected minute-by-minute fresh information on geolocation and system time of neighboring reference GeoClients for the enforcement of usage control. 

\subsection{Threat Model}
In the previous sections we saw the design details that make use of \acp{tee}.
Nevertheless, we must consider that the design is only secure under specific conditions which are detailed in the following sections.
In general, the attack model consists of a malicious host OS or hypervisor at the data consumer according to the definition of Goldreich \cite[Section~7.2.3]{goldreich2009foundations}.
An attacker may intercept or manipulate data being reached into the TEE's execution and may launch or stop the execution.
However, \textbf{we assume that the promised security guarantees of the \acp{tee} are kept} and that applied cryptography is not broken. 
This means that the cryptographic key material which is stored inside of the TEE's CPU can not be leaked and no TEE signatures can be forged and that the TEE's internal memory is confidential.
Therefore, an emulation of an execution inside of a TEE which would allow manipulations is not possible after the TEE's native attestation was validated, as this would require a genuine signature.
We also consider a state-level-attacker for the GeoMesh system, which leaves a vast amount of resources to an attacker.

\subsection{Security Goals}
The protocol design focuses on the following security goals.
(Confidentiality) The data owner's raw data must be kept confidential at all times. 
(Integrity) The integrity of the data owner's raw data must be guaranteed. 
The components of the cloud provider for receiving data and the GeoClients must be integrity-protected and attestable. 
(Availability) The data consumer must be able to process the data provider's raw data at the consumer's machine and a sufficient amount of GeoClients must be available at all times.

\subsection{Security Discussion}

\begin{figure}
  \begin{center}
    \includegraphics[width=\linewidth]{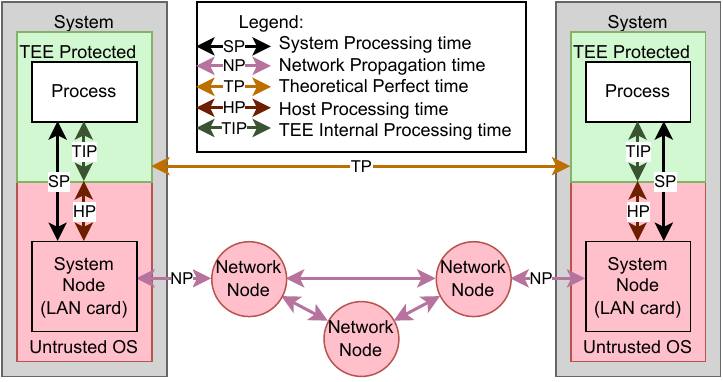}
    \caption{
      Overview of real world and theoretical delays involved during measurement.}
    \label{figure:delays}
  \end{center}
\end{figure}

\textbf{Analysis of Involved Delays:}
Figure \ref{figure:delays} provides an overview of all the processing and network delays which are part of the geolocation attestation.
In this section each delay's role, properties, and possible attack vectors are detailed.
We introduce the term System Processing time (SP blue) to sum up \ac{tee} Internal Processing time (TIP green) and the Host Processing time (HP red).
TIP starts directly at the \ac{tee}-protected Process and ends when the protection realm of the \ac{tee} is left.
HP starts at the outside of the \ac{tee} and ends at the system's network card, whose processing is included.
The Network Propagation time (NP pink) covers all involved end-to-end network nodes, highly depends on the network topology and makes up the biggest overall delay.
NP starts at one network card and ends at another GeoClient's network card.
The last important, required delay is the Theoretical Perfect time (TP orange) which represents a  theoretical perfect connection directly between two \acp{tee} using speed of light as propagation time. 

The \ac{tee} Internal Processing time can be narrowed down, since a \ac{tee}'s attestation report can usually identify the type of CPU which is used.
However, overclocking might let an attacker gain an advantage in the dimension of microseconds, which can be neglected compared to ms of Network Propagation time.
The Host Processing time can in theory also be highly reduced and optimized using specialized hardware. 
Nevertheless, here again only delays in the dimension of microseconds could be gained which can again be neglected for our use case.
Section \ref{evaluation} provides further insights on the evaluation of measured delays.

The Theoretical Perfect time propagating directly from \ac{tee} to \ac{tee} poses the biggest attack vector, because it may highly differ from the Network Propagation time. 
The reason for this difference is the slower-than-light signal propagation in fibre, copper, etc., the greater distance due to network topology, and the processing time required for signal amplification, error correction, routing and other processes that delay data packets.
Another factor is the lower transmission speed in the medium of the involved fiber glass and copper wires compared to theoretical free space.
For instance, if someone was able to create a private network consisting of fiber glass or highly efficient radio transmission systems directly connecting to the referencing GeoClients, the delay could be reduced to aforementioned physical constants, rendering a calibration based on real-world network measurements useless. 
In other words, nothing can prevent an attacker with unlimited resources from establishing a faster direct connection to a single reference node. 
Still, establishing faster connections, which make room for slight inaccuracies of a system's location for an attacker, is computationally infeasible if the attacker does not know which reference nodes will be used in advance, as it is required in \ourtitle{}.

\textbf{\ac{tee} Architecture:} 
Breaking confidentiality and integrity of communication is not possible because the key material between provider and \ac{tee} is established via TLS-tunnels. 
Public parts of the handshake are included in the digitally signed attestation report which is exchanged during the communication startup phase.

The interception or delay of messages by a malicious host or network devices is possible because the \ac{tee} passes on its encrypted traffic to the host OS and through the network.
For geolocation, our approach uses minimum measured delays to derive the minimum distance to other nodes and then calculates the physically possible area from these minimum delays.
Thus, an attacker who is able to delay some measurements can only expand the calculated area, but never move it so that the real location is outside of it. 
Delays cannot be avoided because the network messages pass through several untrusted devices. Also, the exact route cannot be predicted.

In the scope of this work, the Data Provider itself is not placed inside a \ac{tee}, even though it could for additional mutual attestation. 
The Data Consumer takes the role of the prover and is the object which needs to be attested to the Data Provider which is the verifier.
No attestation the other way around is necessary at this point because the Data Provider and its delivered data must be trusted anyways. 

\textbf{Spoofing Issues:} 
Spoofing the \ac{tee}-protected Data Processor and GeoClients to intercept communication is defended via use of \ac{tee}-bound TLS tunnels and \ac{pki} as previously mentioned. 
Spoofing the Data Consumer to receive data from the Data Provider without a subscription using a valid system attested by a \ac{tee} is generally possible. 
Therefore, the Data Consumer must prove the ownership of a valid, previously agreed upon data subscription for receiving the requested sets of data.
Consequently, the data consumer must bootstrap the Data Processor with a secret or some sort of credential during runtime to perform the required authentication.
The resulting report of measurements of the involved GeoClients can not be spoofed as we expect the TEEs to hold their security promises, which means that the digital signature of the TEE can not be forged. 
Additionally, launching a malicious version of a GeoClient, which allows to leak raw data for instance, is also not possible due to the software fingerprinting check of the native attestation.
We expect implementations of GeoClients to be correctly implemented and known beforehand.
Unknown implementations of GeoClients can not participate in \ourtitle{}.

\textbf{Relocation of \acp{tee}:}
For some \ac{tee} technologies, such as AMD \ac{sev}, it is possible to pause execution of a \ac{vm} and export it to another AMD \ac{sev} system, import it there and continue its further execution.
Even though this might seem to be a generally useful feature, its usage would bypass the geolocation mechanism based on \ac{tee} features assumed in our work. 
Therefore, we denote that the \ac{tee}'s configuration must never allow such kind of migration.
This can be technically enforced, since the identity and signature of the CPU performing the current attestation can be linked between different reports.
In case of Intel \ac{sgx}, the corresponding attestation configuration can be set at the Intel Attestation Service.
If binding the CPU would not be possible for some \ac{tee} technology, then repeated random attestations could be applied to detect migration to distant areas and prevent release of data or further computation.

\textbf{Security of GeoMesh:}
The security of the GeoMesh system highly depends on the number of participants for the following reason.
The more participants are available, the less likely it is that an attacker can manipulate all network delays for all requested GeoClients at once.
A similar security concept exists for the TOR-network whose security also increases with higher number of participants \cite{goldschlag1999onion}.
If the system which needs to prove its geolocation could choose its own favorite GeoClients as reference nodes, then adversarial network manipulation on a large scale could be orchestrated.
For this reason, the choice of GeoClients is not in the hand of the proving party, but controlled by the verifying party, which is the Data Provider in our case.

As an additional level of protection, the Data Processor can be designed to not only attest the GeoClients but instead perform a mutual native attestation.
This would mean, that the Data Processor does not just profit from the GeoClients features but itself must act as an active GeoClient.
If all users of the decentralized geolocation mesh were required to become participants themselves, this would encourage higher numbers of participants.
Furthermore, such a requirement would serve as an anti-leeching mechanism and protect the GeoMesh from being overwhelmed with too many requests of non-participating clients.

\textbf{Precision of Delay Measurements:}
In Section \ref{label:sequencediagramsdetailed} within Phase D the Data Provider's check of the geolocation report was marked as optional. 
In theory the attested code base of the Data Processor could immediately enforce given rules.
However, it is impossible for the Data Provider to choose the closest known reference points for an unknown peer from the beginning on.
Therefore, it might be useful for the Data Provider to narrow down the list of given reference GeoClients in an iterative way to provide further accuracy and generally lower measured delays.
Having a documentation of multiple consequent geolocation attestations with increasing precision would generally increase the reliability of the measurements.

\subsection{Use Cases}
The concept described allows a remote peer to process collections of data sets while the provider enforces rules for the consumer's processing. 
The typical data querying use cases which can be addressed using our work are denoted in the following passages.
However, we also want to explain the limitations of our concept. 
It does not aim to cover streams of data, due to the fact that simpler solutions exist to enforce time constraints in this area.
For instance, whenever a data provider wants to end the remote processing of a data stream, the provider simply ends its transmission. 
Therefore, we focus on use cases where the complete dataset is outsourced for processing.

\subsubsection{Time and location restricted licenses}
Licenses may restrict the usage of data to a specific time, location or both. 
One example are media rentals with only a limited time availability in the country where it was bought. 
But also, in the industrial environment these restrictions are prevalent. 
This might be repair instructions for a specific engine only sold in Europe or a set of data sold for different prices in different regions. 
Another example are licenses for sensors as part of collaborations, whereas usage will be restricted to the duration of the collaboration and limited to countries with strong data protection laws and legal security. As well as typical commercial data stores, that want to rent their collections of data only for a certain amount of time for a certain region.
It is important to note, that this can only be enforced until the data leaves the Data Consumer's \ac{tee}. Once processed data is released, no further restrictions can be enforced, but even if the usage control limitations are no longer met, no more data can be retrieved from the \ac{tee}.

\subsubsection{Short-Lived and Regulated Data}
Another example use case involves sets of data which must never leave a country in its raw form and must be available for a short period of time, only.
During elections the processing of data is always limited in time, specifically to the end of the election period.
When it comes to collecting the votes on resources that are owned by the given country, stopping the illegitimate addition of further votes after a deadline has passed can be enforced with our design.
The confidentiality restriction to never pass on data to servers not within the country can also be directly enforced.
The same holds true for projects involving medical data, where patient's consent to share their raw data is only acceptable within their own country's borders for a time interval of their choice.

\section{Implementation}\label{implementation}

This paper aims to focus on enforcing geolocation and time constraints, which is also the implementation's scope.
In order to put the same application into different TEEs we chose to implement the design in a TEE-agnostic way.
Therefore, we used Go programming language version 1.20.0 and the Gramine-framework (formerly known as Graphene-SGX) \cite{tsai2017graphene} for Intel SGX and a Linux-VM for AMD-SEV.
We implemented a simplified system of GeoClients using a single-threaded server and client with the functionalities to measure minimal delays to another as described in our design. 
This way, no additional measurement error from the network path become part of the analysis.
The server will remain unprotected, whereas the client implementation is placed inside the corresponding TEE.

The GeoClients were implemented as UDP based web services. 
The encrypted communication between GeoClients and DataProcessor was established by letting the client randomly create an AES-128-bit secret key with an additional nonce. 
The client encrypted these pieces of information using the GeoClient's public key and the nonce and secret key were used for the following encrypted HTTP communication.
It was not the aim of our implementation to cover all basic functionalities of usage control, because this topic has already been addressed by the scientific community, by work similar to Pretschner et al. \cite{pretschner2006}.
\section{Evaluation}\label{evaluation}

We evaluated our implementation in the two most commonly used \acp{tee} available in cloud environments, Intel SGX and AMD SEV.
To gain insights into the distribution of delays for different numbers of repetitions (UDP packet round-trips), we performed 100 measurements for each test setup.
The test setups were: (1) AMD CPU with SEV, (2) AMD CPU without SEV, (3) Intel CPU with SGX and (4) Intel CPU without SGX.
For Intel SGX and SEV, we benchmarked our setup with 10, 100, 1000 and 10000 repetitions to compensate higher delay variation.

For each of the four setups we determined the maximum range of delays and calculated its impact on geolocation accuracy.
Under idealized transmission conditions, which we assume due to our attacker model, packets travel at the speed of light, i.e. 0.3 kilometres per µs.
Since packets have to travel that distance two times, a distance of 0.15 kilometres per µs round-trip time is a strong upper bound for the physical distance due to general relativity.
The data is summed up in Table \ref{tab:delays}, showing that the resulting inaccuracies induced by TEEs range from 10 to 40 km.

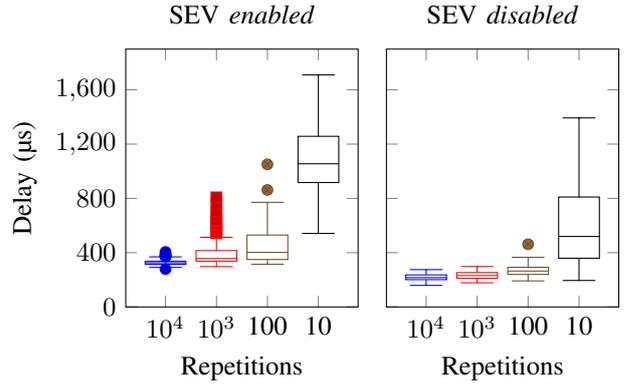
\begin{figure}[t]
    \centering
    \begin{tikzpicture}
            \begin{axis}[
                title={SEV \textit{enabled}},
                xtick={1,2,3,4},
                xticklabels={$10^4$, $10^3$, 100, 10},
                ytick distance=400,
                ymin=0, ymax=1900,
                xlabel=Repetitions, 
                ylabel=Delay (µs),
                boxplot/draw direction=y,width=\columnwidth/2+0.5cm,height=5cm]
            \addplot+[boxplot]
            table[row sep=\\,y index=0] {
                345\\ 345\\ 346\\ 307\\ 330\\ 324\\ 321\\ 325\\ 324\\ 317\\ 363\\ 340\\ 381\\ 303\\ 333\\ 341\\ 292\\ 336\\ 312\\ 321\\ 369\\ 337\\ 326\\ 317\\ 405\\ 335\\ 319\\ 316\\ 311\\ 335\\ 291\\ 359\\ 318\\ 305\\ 314\\ 394\\ 330\\ 317\\ 355\\ 325\\ 327\\ 331\\ 319\\ 317\\ 312\\ 327\\ 358\\ 305\\ 375\\ 298\\ 299\\ 332\\ 347\\ 327\\ 311\\ 298\\ 323\\ 332\\ 301\\ 372\\ 322\\ 325\\ 338\\ 374\\ 331\\ 304\\ 319\\ 330\\ 292\\ 344\\ 330\\ 296\\ 322\\ 344\\ 323\\ 335\\ 331\\ 313\\ 376\\ 319\\ 327\\ 305\\ 309\\ 300\\ 341\\ 302\\ 324\\ 310\\ 336\\ 335\\ 341\\ 315\\ 279\\ 324\\ 332\\ 315\\ 343\\ 336\\ 345\\ 329\\ 
            };
            \addplot+[boxplot]
            table[row sep=\\,y index=0] {
                504\\ 318\\ 342\\ 311\\ 392\\ 387\\ 387\\ 320\\ 324\\ 357\\ 427\\ 365\\ 371\\ 463\\ 613\\ 373\\ 706\\ 394\\ 331\\ 340\\ 409\\ 489\\ 377\\ 323\\ 348\\ 297\\ 314\\ 342\\ 365\\ 378\\ 336\\ 582\\ 367\\ 348\\ 401\\ 561\\ 633\\ 383\\ 451\\ 337\\ 761\\ 379\\ 355\\ 340\\ 362\\ 475\\ 368\\ 348\\ 297\\ 347\\ 357\\ 333\\ 326\\ 354\\ 335\\ 393\\ 675\\ 569\\ 346\\ 513\\ 307\\ 309\\ 334\\ 351\\ 641\\ 347\\ 311\\ 551\\ 369\\ 376\\ 404\\ 604\\ 346\\ 303\\ 338\\ 380\\ 340\\ 485\\ 350\\ 337\\ 381\\ 632\\ 344\\ 350\\ 334\\ 344\\ 306\\ 437\\ 544\\ 411\\ 429\\ 309\\ 350\\ 443\\ 344\\ 344\\ 340\\ 312\\ 808\\ 331\\ 
            };
            \addplot+[boxplot]
            table[row sep=\\,y index=0] {
                679\\ 350\\ 325\\ 403\\ 632\\ 366\\ 593\\ 331\\ 390\\ 525\\ 460\\ 762\\ 345\\ 451\\ 334\\ 354\\ 379\\ 341\\ 493\\ 632\\ 506\\ 642\\ 396\\ 421\\ 347\\ 342\\ 326\\ 726\\ 515\\ 396\\ 696\\ 416\\ 380\\ 351\\ 719\\ 340\\ 439\\ 338\\ 406\\ 358\\ 401\\ 688\\ 367\\ 862\\ 321\\ 446\\ 484\\ 384\\ 335\\ 383\\ 324\\ 613\\ 362\\ 569\\ 412\\ 498\\ 579\\ 360\\ 566\\ 1051\\ 489\\ 369\\ 727\\ 315\\ 333\\ 370\\ 349\\ 411\\ 514\\ 593\\ 350\\ 339\\ 389\\ 321\\ 385\\ 337\\ 512\\ 387\\ 359\\ 482\\ 445\\ 387\\ 335\\ 618\\ 587\\ 521\\ 636\\ 344\\ 325\\ 338\\ 335\\ 544\\ 519\\ 327\\ 459\\ 367\\ 771\\ 478\\ 728\\ 620\\ 
            };
            \addplot+[boxplot]
            table[row sep=\\,y index=0] {
                1214\\ 938\\ 817\\ 1287\\ 943\\ 887\\ 991\\ 900\\ 888\\ 687\\ 901\\ 827\\ 764\\ 1363\\ 863\\ 918\\ 882\\ 854\\ 935\\ 1320\\ 1039\\ 1097\\ 755\\ 1539\\ 840\\ 983\\ 1146\\ 1501\\ 993\\ 1119\\ 1123\\ 1106\\ 819\\ 542\\ 1070\\ 1445\\ 1053\\ 880\\ 1131\\ 1051\\ 1145\\ 937\\ 1271\\ 929\\ 1180\\ 1118\\ 1401\\ 1277\\ 913\\ 1268\\ 923\\ 1082\\ 1432\\ 740\\ 857\\ 1078\\ 1168\\ 1447\\ 1083\\ 1306\\ 863\\ 984\\ 1393\\ 1055\\ 1232\\ 1273\\ 1254\\ 1169\\ 924\\ 1055\\ 993\\ 1139\\ 1231\\ 1530\\ 998\\ 1043\\ 849\\ 930\\ 1298\\ 887\\ 1281\\ 1226\\ 1434\\ 978\\ 1364\\ 1092\\ 776\\ 890\\ 892\\ 927\\ 1182\\ 1271\\ 1281\\ 1289\\ 1035\\ 954\\ 1081\\ 1328\\ 1710\\ 1048\\ 
            };
            \end{axis}
        \end{tikzpicture}
    \begin{tikzpicture}
        \centering
            \begin{axis}[
                title={SEV \textit{disabled}},
                xtick={1,2,3,4},
                xticklabels={$10^4$, $10^3$, 100, 10},
                ytick distance=400,
                ymin=0, ymax=1900,
                xlabel=Repetitions, 
                scaled y ticks=false,
                yticklabels=\empty,
                boxplot/draw direction=y,width=\columnwidth/2+0.5cm,height=5cm]
            \addplot+[boxplot]
            table[row sep=\\,y index=0] {
                212\\ 186\\ 232\\ 201\\ 172\\ 204\\ 179\\ 232\\ 249\\ 194\\ 242\\ 175\\ 200\\ 211\\ 219\\ 193\\ 205\\ 227\\ 159\\ 216\\ 228\\ 210\\ 236\\ 204\\ 266\\ 248\\ 191\\ 213\\ 220\\ 222\\ 219\\ 165\\ 215\\ 218\\ 233\\ 218\\ 202\\ 259\\ 232\\ 219\\ 236\\ 211\\ 232\\ 209\\ 258\\ 250\\ 206\\ 201\\ 221\\ 170\\ 216\\ 190\\ 250\\ 265\\ 219\\ 250\\ 247\\ 249\\ 174\\ 190\\ 181\\ 174\\ 176\\ 174\\ 180\\ 239\\ 217\\ 211\\ 256\\ 236\\ 275\\ 215\\ 224\\ 187\\ 249\\ 210\\ 217\\ 209\\ 206\\ 213\\ 257\\ 219\\ 255\\ 231\\ 270\\ 188\\ 200\\ 225\\ 216\\ 224\\ 205\\ 213\\ 208\\ 253\\ 273\\ 211\\ 256\\ 170\\ 256\\ 175\\   
            };
            \addplot+[boxplot]
            table[row sep=\\,y index=0] {
                283\\ 261\\ 274\\ 232\\ 249\\ 294\\ 229\\ 180\\ 271\\ 262\\ 220\\ 201\\ 214\\ 225\\ 225\\ 222\\ 226\\ 248\\ 236\\ 196\\ 280\\ 267\\ 206\\ 214\\ 220\\ 286\\ 229\\ 273\\ 235\\ 191\\ 217\\ 240\\ 268\\ 218\\ 238\\ 298\\ 284\\ 274\\ 192\\ 280\\ 220\\ 277\\ 240\\ 182\\ 212\\ 209\\ 262\\ 254\\ 205\\ 253\\ 196\\ 198\\ 229\\ 181\\ 220\\ 208\\ 222\\ 269\\ 202\\ 214\\ 190\\ 237\\ 232\\ 201\\ 245\\ 274\\ 244\\ 238\\ 241\\ 267\\ 247\\ 185\\ 226\\ 205\\ 215\\ 272\\ 207\\ 266\\ 224\\ 232\\ 223\\ 218\\ 189\\ 178\\ 241\\ 203\\ 261\\ 250\\ 238\\ 183\\ 278\\ 229\\ 232\\ 235\\ 199\\ 248\\ 233\\ 248\\ 200\\ 228\\  
            };
            \addplot+[boxplot]
            table[row sep=\\,y index=0] {
                243\\ 279\\ 237\\ 300\\ 278\\ 196\\ 224\\ 253\\ 267\\ 237\\ 261\\ 245\\ 252\\ 240\\ 262\\ 270\\ 237\\ 204\\ 284\\ 297\\ 228\\ 272\\ 285\\ 255\\ 264\\ 246\\ 239\\ 305\\ 214\\ 305\\ 236\\ 275\\ 199\\ 255\\ 260\\ 231\\ 276\\ 255\\ 233\\ 250\\ 254\\ 265\\ 285\\ 287\\ 271\\ 280\\ 298\\ 220\\ 318\\ 275\\ 214\\ 292\\ 277\\ 238\\ 349\\ 282\\ 301\\ 247\\ 208\\ 333\\ 237\\ 298\\ 255\\ 254\\ 203\\ 276\\ 296\\ 237\\ 286\\ 261\\ 305\\ 252\\ 192\\ 197\\ 299\\ 253\\ 462\\ 252\\ 274\\ 238\\ 302\\ 305\\ 251\\ 293\\ 272\\ 267\\ 296\\ 300\\ 255\\ 284\\ 304\\ 280\\ 203\\ 366\\ 318\\ 293\\ 300\\ 227\\ 318\\ 244\\  
            };
            \addplot+[boxplot]
            table[row sep=\\,y index=0] {
                425\\ 877\\ 337\\ 302\\ 968\\ 713\\ 825\\ 375\\ 399\\ 512\\ 1248\\ 548\\ 287\\ 738\\ 365\\ 507\\ 952\\ 288\\ 1141\\ 582\\ 710\\ 600\\ 638\\ 372\\ 857\\ 883\\ 894\\ 920\\ 838\\ 676\\ 282\\ 999\\ 805\\ 925\\ 195\\ 373\\ 350\\ 431\\ 384\\ 349\\ 882\\ 302\\ 761\\ 344\\ 854\\ 1120\\ 632\\ 267\\ 594\\ 362\\ 334\\ 672\\ 748\\ 798\\ 520\\ 383\\ 1205\\ 293\\ 305\\ 347\\ 674\\ 967\\ 881\\ 329\\ 371\\ 439\\ 564\\ 291\\ 370\\ 1044\\ 639\\ 367\\ 449\\ 290\\ 753\\ 339\\ 365\\ 382\\ 270\\ 345\\ 534\\ 1175\\ 420\\ 592\\ 316\\ 519\\ 271\\ 461\\ 785\\ 362\\ 1393\\ 317\\ 836\\ 433\\ 450\\ 292\\ 521\\ 1011\\ 604\\ 1142\\ 
            };
            \end{axis}
        \end{tikzpicture}
    \caption{Delay measurements using AMD with SEV enabled (left) and with SEV disabled (right). 
        No significant impact was measured, whether SEV was activated or not.
        With higher numbers of repetitions for each of 100 test sequences the variance of measured lowest delays generally decreases, but almost never the overall minimum delay.}
    \label{figure:delayamd}
\end{figure}

\begin{figure}[t]
    \centering
    \begin{tikzpicture}
            \begin{axis}[
                title={SGX \textit{enabled}},
                xtick={1,2,3,4},
                xticklabels={$10^4$, $10^3$, 100, 10},
                ytick distance=100,
                ymin=0, ymax=400,
                xlabel=Repetitions, 
                ylabel=Delay (µs),
                boxplot/draw direction=y,width=\columnwidth/2+0.5cm,height=5cm]
            \addplot+[boxplot]
            table[row sep=\\,y index=0] {
77\\ 76\\ 77\\ 76\\ 74\\ 78\\ 77\\ 76\\ 76\\ 76\\ 75\\ 77\\ 76\\ 76\\ 77\\ 75\\ 75\\ 73\\ 77\\ 74\\ 74\\ 75\\ 77\\ 74\\ 76\\ 76\\ 77\\ 76\\ 76\\ 77\\ 77\\ 73\\ 75\\ 77\\ 72\\ 75\\ 75\\ 78\\ 75\\ 77\\ 77\\ 77\\ 74\\ 76\\ 77\\ 76\\ 78\\ 75\\ 77\\ 76\\ 76\\ 77\\ 74\\ 78\\ 76\\ 77\\ 74\\ 76\\ 78\\ 77\\ 76\\ 78\\ 78\\ 80\\ 76\\ 75\\ 78\\ 77\\ 75\\ 73\\ 76\\ 76\\ 75\\ 77\\ 76\\ 76\\ 71\\ 76\\ 76\\ 75\\ 76\\ 76\\ 75\\ 76\\ 76\\ 75\\ 76\\ 75\\ 75\\ 76\\ 75\\ 76\\ 76\\ 74\\ 77\\ 76\\ 77\\ 77\\ 78\\ 73\\ 
            };
            \addplot+[boxplot]
            table[row sep=\\,y index=0] {
79\\ 82\\ 81\\ 75\\ 78\\ 84\\ 77\\ 79\\ 80\\ 77\\ 79\\ 78\\ 77\\ 79\\ 78\\ 80\\ 75\\ 81\\ 80\\ 77\\ 77\\ 80\\ 82\\ 80\\ 78\\ 78\\ 78\\ 81\\ 82\\ 78\\ 79\\ 79\\ 83\\ 78\\ 78\\ 76\\ 76\\ 76\\ 77\\ 76\\ 78\\ 79\\ 78\\ 81\\ 78\\ 78\\ 76\\ 79\\ 78\\ 83\\ 81\\ 78\\ 81\\ 79\\ 77\\ 76\\ 78\\ 80\\ 77\\ 81\\ 83\\ 79\\ 80\\ 80\\ 76\\ 79\\ 79\\ 77\\ 79\\ 79\\ 78\\ 77\\ 80\\ 76\\ 82\\ 80\\ 78\\ 75\\ 80\\ 77\\ 81\\ 78\\ 78\\ 77\\ 78\\ 77\\ 80\\ 76\\ 78\\ 79\\ 73\\ 82\\ 78\\ 76\\ 79\\ 79\\ 77\\ 76\\ 78\\ 81\\ 
            };
            \addplot+[boxplot]
            table[row sep=\\,y index=0] {
78\\ 87\\ 78\\ 74\\ 81\\ 79\\ 78\\ 78\\ 79\\ 77\\ 77\\ 85\\ 134\\ 79\\ 84\\ 125\\ 126\\ 80\\ 237\\ 78\\ 82\\ 102\\ 77\\ 78\\ 187\\ 82\\ 117\\ 84\\ 81\\ 76\\ 136\\ 98\\ 82\\ 78\\ 173\\ 82\\ 78\\ 87\\ 84\\ 80\\ 83\\ 79\\ 84\\ 78\\ 83\\ 86\\ 88\\ 77\\ 179\\ 84\\ 122\\ 91\\ 77\\ 79\\ 117\\ 164\\ 87\\ 83\\ 79\\ 78\\ 83\\ 84\\ 83\\ 82\\ 248\\ 78\\ 296\\ 84\\ 127\\ 80\\ 81\\ 86\\ 153\\ 76\\ 85\\ 82\\ 81\\ 110\\ 97\\ 83\\ 87\\ 123\\ 85\\ 113\\ 79\\ 79\\ 127\\ 76\\ 113\\ 83\\ 94\\ 82\\ 84\\ 83\\ 79\\ 84\\ 77\\ 80\\ 86\\ 79\\ 
            };
            \addplot+[boxplot]
            table[row sep=\\,y index=0] {
86\\ 123\\ 209\\ 107\\ 128\\ 115\\ 118\\ 205\\ 127\\ 116\\ 124\\ 188\\ 123\\ 125\\ 110\\ 155\\ 139\\ 98\\ 101\\ 145\\ 83\\ 117\\ 181\\ 167\\ 121\\ 153\\ 104\\ 124\\ 121\\ 159\\ 117\\ 93\\ 119\\ 86\\ 104\\ 157\\ 116\\ 237\\ 203\\ 212\\ 188\\ 114\\ 137\\ 149\\ 111\\ 330\\ 138\\ 148\\ 83\\ 146\\ 92\\ 118\\ 114\\ 186\\ 125\\ 137\\ 132\\ 229\\ 165\\ 107\\ 128\\ 139\\ 116\\ 77\\ 156\\ 152\\ 267\\ 195\\ 183\\ 134\\ 140\\ 261\\ 244\\ 124\\ 150\\ 162\\ 202\\ 127\\ 169\\ 161\\ 197\\ 139\\ 141\\ 193\\ 149\\ 150\\ 139\\ 118\\ 149\\ 129\\ 222\\ 180\\ 84\\ 161\\ 206\\ 158\\ 119\\ 111\\ 131\\ 164\\ 
            };
            \end{axis}
        \end{tikzpicture}
    \begin{tikzpicture}
        \centering 
            \begin{axis}[
                title={SGX \textit{disabled}},
                xtick={1,2,3,4},
                xticklabels={$10^4$, $10^3$, 100, 10},
                ytick distance=100,
                ymin=0, ymax=400,
                xlabel=Repetitions, 
                scaled y ticks=false,
                yticklabels=\empty,
                boxplot/draw direction=y,width=\columnwidth/2+0.5cm,height=5cm]
            \addplot+[boxplot]
            table[row sep=\\,y index=0] {
29\\ 28\\ 28\\ 29\\ 28\\ 27\\ 28\\ 28\\ 28\\ 28\\ 28\\ 28\\ 28\\ 28\\ 28\\ 28\\ 28\\ 28\\ 28\\ 28\\ 28\\ 28\\ 28\\ 28\\ 28\\ 28\\ 28\\ 28\\ 28\\ 28\\ 28\\ 28\\ 28\\ 28\\ 28\\ 28\\ 28\\ 27\\ 28\\ 28\\ 28\\ 28\\ 28\\ 28\\ 28\\ 28\\ 28\\ 28\\ 28\\ 28\\ 28\\ 28\\ 28\\ 28\\ 28\\ 27\\ 28\\ 28\\ 28\\ 28\\ 28\\ 28\\ 28\\ 28\\ 28\\ 28\\ 27\\ 28\\ 27\\ 28\\ 28\\ 28\\ 28\\ 28\\ 28\\ 28\\ 28\\ 28\\ 28\\ 28\\ 27\\ 28\\ 28\\ 28\\ 28\\ 28\\ 27\\ 28\\ 28\\ 29\\ 29\\ 28\\ 28\\ 28\\ 28\\ 27\\ 28\\ 28\\ 28\\ 28\\ 
            };
            \addplot+[boxplot]
            table[row sep=\\,y index=0] {
29\\ 29\\ 29\\ 30\\ 30\\ 30\\ 29\\ 29\\ 29\\ 29\\ 29\\ 30\\ 30\\ 29\\ 28\\ 28\\ 29\\ 29\\ 29\\ 29\\ 30\\ 28\\ 30\\ 29\\ 29\\ 29\\ 29\\ 30\\ 30\\ 29\\ 29\\ 29\\ 30\\ 29\\ 29\\ 30\\ 29\\ 29\\ 29\\ 28\\ 29\\ 29\\ 29\\ 30\\ 28\\ 29\\ 30\\ 29\\ 29\\ 28\\ 29\\ 30\\ 30\\ 30\\ 30\\ 28\\ 29\\ 29\\ 29\\ 29\\ 29\\ 28\\ 29\\ 30\\ 30\\ 28\\ 29\\ 28\\ 29\\ 29\\ 30\\ 30\\ 30\\ 30\\ 30\\ 28\\ 29\\ 28\\ 29\\ 29\\ 30\\ 29\\ 29\\ 29\\ 29\\ 29\\ 29\\ 29\\ 30\\ 29\\ 28\\ 29\\ 28\\ 28\\ 50\\ 30\\ 29\\ 28\\ 29\\ 30\\ 
            };
            \addplot+[boxplot]
            table[row sep=\\,y index=0] {
31\\ 36\\ 35\\ 38\\ 29\\ 35\\ 35\\ 32\\ 31\\ 35\\ 32\\ 30\\ 40\\ 33\\ 37\\ 35\\ 33\\ 36\\ 34\\ 51\\ 35\\ 34\\ 34\\ 33\\ 30\\ 77\\ 30\\ 33\\ 30\\ 31\\ 35\\ 33\\ 34\\ 33\\ 38\\ 32\\ 31\\ 28\\ 58\\ 30\\ 34\\ 31\\ 31\\ 33\\ 33\\ 30\\ 32\\ 30\\ 36\\ 34\\ 30\\ 35\\ 33\\ 32\\ 29\\ 29\\ 31\\ 33\\ 44\\ 33\\ 33\\ 32\\ 32\\ 29\\ 34\\ 33\\ 32\\ 36\\ 34\\ 34\\ 36\\ 35\\ 30\\ 31\\ 36\\ 50\\ 41\\ 30\\ 29\\ 32\\ 32\\ 32\\ 29\\ 32\\ 35\\ 32\\ 31\\ 32\\ 30\\ 33\\ 36\\ 31\\ 29\\ 34\\ 34\\ 35\\ 35\\ 35\\ 29\\ 29\\ 
            };
            \addplot+[boxplot]
            table[row sep=\\,y index=0] {
86\\ 38\\ 79\\ 124\\ 32\\ 94\\ 76\\ 71\\ 73\\ 46\\ 70\\ 90\\ 128\\ 112\\ 165\\ 60\\ 112\\ 112\\ 144\\ 55\\ 110\\ 48\\ 36\\ 62\\ 104\\ 59\\ 108\\ 143\\ 59\\ 75\\ 57\\ 54\\ 37\\ 52\\ 46\\ 36\\ 42\\ 53\\ 114\\ 146\\ 92\\ 40\\ 42\\ 97\\ 83\\ 54\\ 38\\ 68\\ 66\\ 100\\ 88\\ 52\\ 55\\ 61\\ 158\\ 60\\ 100\\ 72\\ 148\\ 71\\ 31\\ 307\\ 131\\ 54\\ 91\\ 47\\ 39\\ 102\\ 48\\ 77\\ 108\\ 128\\ 76\\ 120\\ 100\\ 46\\ 62\\ 55\\ 61\\ 42\\ 59\\ 63\\ 36\\ 89\\ 78\\ 89\\ 86\\ 48\\ 55\\ 93\\ 63\\ 78\\ 40\\ 49\\ 66\\ 42\\ 68\\ 48\\ 124\\ 66\\ 
            };
            \end{axis}
        \end{tikzpicture}
    \caption{Delay measurements using Intel with SGX enabled (left) and with SGX disabled (right). An impact was measured, whether SGX was activated or not. With higher numbers of repetitions for each of 100 test sequences the variance of measured lowest delays generally decreases, but not the overall minimum delay.}
    \label{figure:delaysgx}
\end{figure}
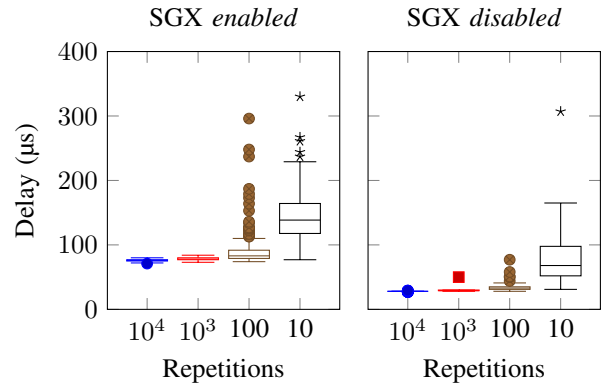

The distributions of minimum delay for AMD SEV are provided in Figure \ref{figure:delayamd} and for Intel SGX in Figure \ref{figure:delaysgx}.
In these graphs the most extreme values, lower and upper quartile and median measurements are plotted for each of the setups. 
For both systems, a clear increase in overall delay and also a higher variance between minimum and maximum delays could be observed.
For the AMD setup, we observed that the \ac{tee} significantly increases the variance of measurements and produces many more outliers compared to the reference VM. The minimal delay measured is also significantly higher, with 279 µs vs. 159 µs, respectively.
For Intel SGX, the \ac{tee} has a major impact on performance, almost doubling the overall delays, though keeping a low distribution.

For the SEV measurements the prototype was executed within two VMs on Google Cloud, one protected using AMD SEV, the other one as a standard VM.
Both VMs have identical hardware specifications (n2d-standard-2 instance, AMD EPYC (Rome), 2 vCPUs, 8 GB vRAM).
Only the boot images, Ubuntu 20.04 LTS Pro FIPS Server and Ubuntu 20.04 LTS, respectively, differ slightly for technical reasons.

For the SGX measurements the prototype was executed and its performance measured on a NUC7PJYH.
The system used was an Intel Pentium Silver J5005 CPU @ 1.50GHz with 8GB of RAM, Linux version Ubuntu 22.04 and kernel version GNU/Linux 4.15.0-159-generic x86\_64.
Additionally, the Gramine-SGX framework version 1.4 was used to execute the go application without having to implement SGX-specific adaptations.

\begin{table*}[t]
\centering
\caption{Delay Ranges, Resulting Range Accuracy and Minimum Delays Measured for and AMD SEV and Intel SGX}
\label{tab:delays}
\begin{tabular}{lllllllllllll}
\rowcolor[HTML]{EFEFEF} 
 & Repetitions & 10000 & 1000 & 100 & 10 &  &  & \cellcolor[HTML]{EFEFEF}Repetitions & \cellcolor[HTML]{EFEFEF}10000 & \cellcolor[HTML]{EFEFEF}1000 & \cellcolor[HTML]{EFEFEF}100 & \cellcolor[HTML]{EFEFEF}10 \\
 & Max Delay (µs) & 405 & 808 & 1051 & 1710 &  & \cellcolor[HTML]{FFFFFF} & \cellcolor[HTML]{FFFFFF}Max Delay (µs) & \cellcolor[HTML]{FFFFFF}80 & \cellcolor[HTML]{FFFFFF}84 & \cellcolor[HTML]{FFFFFF}296 & \cellcolor[HTML]{FFFFFF}330 \\
\multirow{-2}{*}{\begin{tabular}[c]{@{}l@{}}AMD\\ SEV\end{tabular}} & Distance (km) & 60.75 & 121.2 & 157.65 & 256.5 &  & \multirow{-2}{*}{\cellcolor[HTML]{FFFFFF}\begin{tabular}[c]{@{}l@{}}Intel\\ SGX\end{tabular}} & \cellcolor[HTML]{FFFFFF}Distance (km) & \cellcolor[HTML]{FFFFFF}12 & \cellcolor[HTML]{FFFFFF}12.6 & \cellcolor[HTML]{FFFFFF}44.4 & \cellcolor[HTML]{FFFFFF}49.5 \\
\rowcolor[HTML]{EFEFEF} 
\cellcolor[HTML]{EFEFEF} & Max Delay (µs) & 275 & 298 & 462 & 1393 &  & \cellcolor[HTML]{EFEFEF} & Max Delay (µs) & 29 & 50 & 77 & 307 \\
\rowcolor[HTML]{EFEFEF} 
\multirow{-2}{*}{\cellcolor[HTML]{EFEFEF}\begin{tabular}[c]{@{}l@{}}AMD\\ default\end{tabular}} & Distance (km) & 41.25 & 44.7 & 69.3 & 208.95 & \cellcolor[HTML]{EFEFEF} & \multirow{-2}{*}{\cellcolor[HTML]{EFEFEF}\begin{tabular}[c]{@{}l@{}}Intel\\ default\end{tabular}} & Distance (km) & 4.35 & 7.5 & 11.55 & 45.05
\end{tabular}%
\end{table*}

\section{Discussion}\label{discussion}

In the previous sections we explained our derived measurements using our implementation and testing setup. 
The datasets lead to the following interpretations and discussions.

\subsection{Geolocation Precision}
A general limitation is at hand considering the fact that the network delay based geolocation only has a precision level in the dimension of tens or hundreds of km of range \cite{katz2006towards}.
Therefore, it is not possible to precisely attest that a server is within a certain building, or a certain city.
Still, a non-forgeable attested geolocation for average-sized countries or wider areas is indeed possible.
Moving usage control applications inside of TEEs will have an inaccuracy in geolocation, which is acceptable for enforcing regulations, such as the GDPR which demands computing within certain countries. 

As can be seen in Table \ref{tab:delays}, activating the TEE has an additional delay compared to the same system having the TEE deactivated. 
For SGX the additional inaccuracy for our worst case delay with repetitions of 10000 is less than 8 km (12 - 4.35 km) per TEE which means less than 16 km for SGX on both ends of the connection.
For AMD SEV the worst case additional delay is less than 19.5 km (60.75 - 41.25 km) per TEE which means less than 39 km for SEV on both ends.
This result shows, that when our application is launched inside the TEEs testbed we have an additional worst case inaccuracy which downgrades the precision by an additional few dozens of km to the existing state-of-the-art of roughly 67 to 228 km of precision range by Katz-Bessett et al. \cite{katz2006towards}.
Therefore, in comparison to GPS-based geolocation the measured position is less precise but cannot be spoofed if the accuracy mentioned is acceptable for the use case.

\subsection{Field Test in a Cloud Environment}
We additionally tested our prototype in a cloud environment using our AMD image without SEV and measured the delay from a datacenter located in Paris to systems on: 
(A) a neighboring datacenter in the same cluster in Paris, 
(B) a datacenter located in Madrid (1052 km distance), 
and (C) a datacenter located in Northern Virginia USA (roughly 6338 km distance).
For case A we measured roughly 0.668 ms, for case B roughly 18.5 ms, and for case C roughly 76.6 ms. 
All these values come very close to ordinary distance measurement tools, such as the ping command for linux OS, for instance. 

For comparison, 1 ms of network delay corresponds to 300 km of distance at the speed of light. 
Therefore, a round-trip delay of 1 ms corresponds to a distance of 150 km.
Since even a direct neighboring building with 0.668 ms would mean an inaccuracy of more than 75 km, our concept is rather beneficial for long-range distances and does not provide much benefit for very low distances.
However, this is a typical issue for any geolocation system based on network delay.
Nevertheless, it can be partly compensated by the use of a massive amount of GeoClients as reference nodes.

\subsection{Impact of Repetitions}
According to our numbers, larger sets of repeated measurements do typically increase precision significantly. 
Despite higher numbers of repetitions, such as 10000 within one measurement, there seems to a be an early saturation of precision improvement in our setup at 1000 repetitions for the modes SGX activated, SGX deactivated, and AMD SEV deactivated. 
However, for AMD SEV activated still numerous outliers can be seen at 1000 repetitions, and satisfying results can be seen at 10000 repetitions.
In general for AMD, more variance can be observed due to the overhead of having to handle complete instances of virtual machines instead of just processes with context switches, as for Intel.

\subsection{Time Constraints for Large Sets of Outputs}
When it comes to data processing with a single output, a single update iteration with the \ac{tee}-internal system time is required.
However, when it comes to processing sequences of a large number of outputs, repeating the decentralized check of internal system time a vast amount of repetitions would cause a too high performance overhead.
For this reason the TEE-protected data processor must be able to preprocess results, queue them up and release them using a single update iteration.

\section{Conclusion}\label{conclusion}

We developed a TEE-agnostic protocol for enforcing geolocation and time-constraints in the context of usage control.
To enforce these constraints, we make use of an IT infrastructure, which uses a decentralized mesh of GeoClients as reference nodes for geolocation based on network delay.
Our analysis shows that moving existing approaches for geolocation into TEEs has an additional inaccuracy of less than 16km for Intel SGX and less than 39km AMD SEV while at the same time increasing security guarantees.
We strongly encourage the use of network-based delay mechanisms in cases where tens of km of precision are sufficient, instead of relying on spoofable techniques, such as GPS or IP-based ones.

Using our approach it is possible to additionally equip TEE protected usage control systems with constraints for time and geolocation limitations.
This can serve as an additional level of security for cases where previously only legal-based solutions pinned down the location of cloud infrastructures or server farms. 
Attacks, which manipulate the internal processing times of TEEs can additionally be protected using our protocol.
This can increase the security of constraints for trusted computations in cloud environments.
\section*{Acknowledgements}

This work has been funded by the Fraunhofer-Cluster of Excellence ``Cognitive Internet Technologies''.

\bibliographystyle{plain}
\bibliography{paper}

\end{document}